# Understanding modes of negative differential resistance in amorphous and polycrystalline vanadium oxides

Sanjoy Kumar Nandi*[a], Sujan Kumar Das[a,1], Caleb Estherby[b], Angus Gentle[b], Robert G. Elliman[a]

[a]Department of Electronic Materials Engineering, Research School of Physics, Australian National University, Canberra, ACT 2601, Australia
[b]School of Mathematical and Physical Sciences, University of Technology Sydney, NSW 2007, Australia
*E-mail: sanjoy.nandi@anu.edu.au

## Abstract

Metal-oxide-metal devices based on amorphous $VO_x$ are shown to exhibit one of two distinct negative differential resistance (NDR) characteristics depending on the maximum current employed for electroforming. For low compliance currents they exhibit a smooth S-type characteristic and have a temperature-dependent device resistance characterised by an activation energy of 0.25 eV, consistent with conduction in polycrystalline $VO_2$, while for high-compliance currents they exhibit an abrupt snap-back characteristic and a resistance characterised by an activation energy of 0.025 eV, consistent with conduction in oxygen deficient $VO_x$. In both cases, the temperature dependence of the switching voltage implies that the conductivity change is due to the insulator-metal transition in $VO_2$. From this analysis it is concluded that electroforming at low currents creates a conductive filament comprised largely of polycrystalline $VO_2$, while electroforming at high currents creates a composite structure comprised of $VO_2$ and a conductive halo of oxygen deficient $VO_x$. The effect of electroforming on the NDR mode is then explained with reference to a lumped element model of filamentary conduction that includes the effect of a parallel resistance created by the halo. These results provide new insight into the NDR response of vanadium-oxide-based devices and a basis for designing devices with specific characteristics.

## 1. Introduction

Current controlled negative differential resistance (NDR) in metal-oxide-metal devices is of interest as the basis of nanoscale relaxation oscillators for use as solid-state neurons in neuromorphic computing arrays.[1,2] The as-fabricated devices are generally in a high resistance state and require a one-off electroforming step to initiate the NDR response.[3,4] This is typically achieved by subjecting the film to a voltage or current stress sufficient to form a filamentary conduction path through the film (i.e. soft dielectric breakdown), a process mediated by the generation, drift and diffusion of atoms and ions in response to the applied electric field and local Joule heatin.[5, 6] The size, resistance and stability of the resulting filaments depend critically on the forming conditions, and particularly on the maximum forming current and the associated temperature rise caused by Joule heating.[7, 8] The high temperatures associated with electroforming can also cause crystallisation of amorphous films and compositional or structural changes at the oxide/electrode interface that affect the final state of the electroformed device and its switching characteristics.[9, 10] As a consequence, understanding details of the electroforming process is an essential requirement for developing devices with specific characteristics.

---

[1] The permanent address of Sujan Kumar Das is Department of Physics, Jahangirnagar University, Bangladesh.

In amorphous vanadium-oxide based devices the NDR response is generally attributed to the insulator-metal transition in $VO_2$ on the assumption that this phase is crystallised within the filamentary conduction path during electroforming.[9] Such devices generally exhibit smooth S-type NDR due to the heterogeneous nature of the IMT transition and the evolution of the temperature distribution during current-controlled testing[11]. However, they can also exhibit an abrupt snap-back characteristic under certain condition, similar to that observed in $NbO_x$ based devices.[7, 12] This novel NDR mode has the potential to offer new device functionality but its origin continues to be debated.[12, 13]

In this study, we show that electroforming can be used to control the NDR characteristics of amorphous $VO_x$ films and that the snap-back response can be understood from the filament microstructure and its impact on the effective circuit of the device.

## 2. Experimental Details

Two device structures were employed for these studies: metal-oxide-metal (MOM) capacitor structures fabricated with a common bottom electrode (BE) and top electrodes (TE) of 100μm diameter circles defined by a shadow mask; and cross-point devices (2μm x 2μm and 5μm x 5μm ) fabricated using step-by-step photolithography, as shown in Figure 1.[14] In both cases, the devices were fabricated on thermally oxidized (100nm $SiO_2$) Si wafers by sequential layer deposition. The bottom electrodes consisted of a 10 nm-thick Ti adhesion layer and a 50 nm-thick Pt contact layer deposited by sequential e-beam evaporation. A 70 nm thick functional oxide layer of either amorphous $VO_x$ or polycrystalline $VO_2$ was then deposited by reactive sputter deposition from a V target using an $O_2$/Ar ambient maintained at a pressure of 2.3 (or 1.5) mTorr using Ar/$O_2$ flow rates of 58/2 (or 58/10) sccm. Amorphous $VO_x$ films (a-$VO_x$) were achieved by maintaining the substrates at room temperature and polycrystalline $VO_2$ films (pc-$VO_2$) were achieved by post-annealing the film at 450 °C in in a partial vacuum (1.5 Torr air). The devices were completed by adding top electrodes consisting of a 5 nm-thick Ti layer and a 25 nm Pt layer.

The as-deposited oxide films were characterised by grazing incident angle X-ray diffraction (GI-XRD), atomic force microscopy (AFM), Raman spectroscopy (RS) and electron Rutherford backscattering spectrometry (eRBS).[15] Electrical measurements were performed with an Agilent B1500A semiconductor parameter analyser attached to a Signatone probe station (S-1160) and were undertaken in air by applying voltages to top electrode while grounding the bottom electrode.

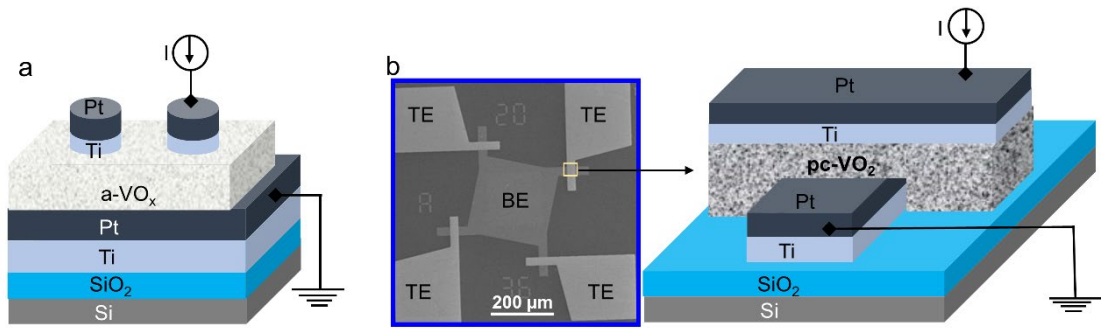

Figure 1: (a) Schematic showing the test structure of Pt/Ti/a-$VO_x$/Pt devices and (b) scanning electron microscopy image of Pt/Ti/pc-$VO_2$/Pt cross-point device showing four 20 μm x 20 μm devices with a common bottom electrode with 3D schematic of the cross-point area.

## 3. Experimental Results and Discussion

### 3.1 Composition and Structure of Films

Figure 2 shows GI-XRD spectra and AFM images of the as-deposited a-$VO_x$ and pc-$VO_2$ films, together with Raman spectra from the pc-$VO_2$ film as a function of temperature. The GI-XRD spectrum from the pc-$VO_2$ film has peaks corresponding to (011), (220), and (022) planes of monoclinic $VO_2$, while that from a-$VO_x$ film is essentially featureless, consistent with the film being amorphous. (The only diffraction peaks observed in this case are from the underlying Pt substrate). Temperature dependent Raman analysis showed that the pc-$VO_2$ film underwent a thermally induced phase transition at temperatures between 40°C and 80°C, consistent with the well-known insulator-metal phase transition in $VO_2$. The surface morphology and roughness of the films was determined from AFM images and was similar for both films, with the RMS roughness measured to be 2.1 and 2.3 nm for the a-VO$_x$ and pc-VO$_2$ films, respectively. These results, combined with eRBS analysis show that the pc-VO$_2$ films are polycrystalline and composed of monoclinic VO$_2$, while the VO$_x$ films are amorphous and have a composition close to V$_2$O$_5$ (i.e. x~2.5).

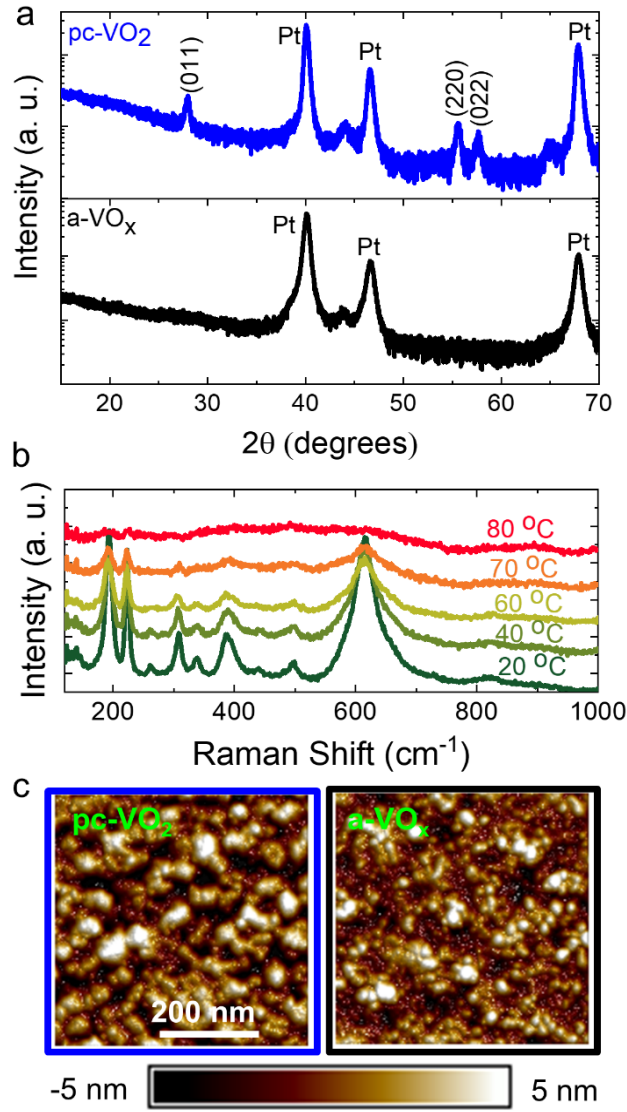

Figure 2: (a) GIXRD of pc-VO$_2$ and a-VO$_x$ films deposited on Pt, (b) Raman spectra from pc-VO$_2$ as a function of temperature, and (c) AFM images of the pc-VO$_2$ and a-VO$_x$ films.

*3.2 Electrical characterization*

As-fabricated devices were highly resistive, with resistances of order several MΩ, and required a one-off electroforming step to initiate threshold switching, as shown in Figure 3. This was achieved by scanning either the voltage or the current and detecting the abrupt conductivity change indicative of filament formation, and typically reduced the device resistance by around an order of magnitude, consistent with the creation of a filamentary conduction path in the oxide layer. For voltage controlled electroforming the maximum current ($I_{CC}$) was limited to avoid device damage. Immediately following electroforming the devices exhibited symmetric threshold switching under voltage controlled testing, with threshold voltages in the range from ±0.45 v to ±2.2 V. Similar behavior has previously been reported in both lateral and vertical device structures and is generally attributed to the thermally induced IMT in $VO_2$ and the associated positive feedback created by Joule heating.[16-18] In the case of a-$VO_x$ devices, this is predicated on the assumption that the $VO_2$ phase is crystallized within the amorphous film during electroforming[9].

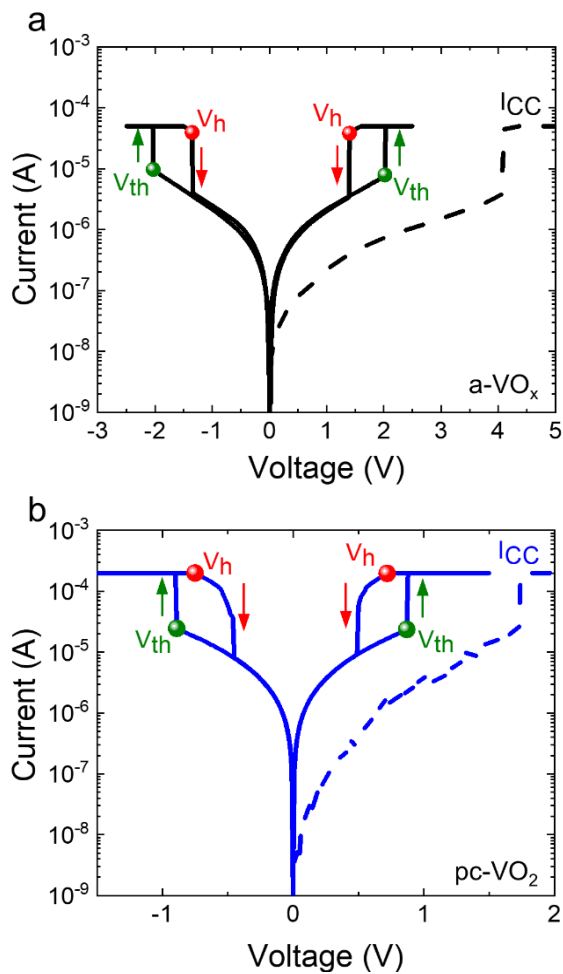

Figure 3: Electroforming (dashed line) and subsequent threshold switching characteristics (solid line) for (a) an a-$VO_x$ capacitor structure (100 μm diameter) and (b) a pc-$VO_2$ cross-point device (2μm x 2 μm).

Figure 4a shows corresponding current-controlled I-V characteristics for these devices. In this case, the current is constrained by the measurement system so that the increase in conductivity is self-limiting and the I-V characteristics vary continuously and the regions of NDR reflect the fact that the conductivity increases superlinearly with current (temperature). [19] Given the significant difference in the initial film properties and device structures, the characteristics of the pc-VO$_2$ and a-VO$_x$ devices are remarkably similar, reflecting the filamentary nature of the conduction process and the common origin of the conductivity change.

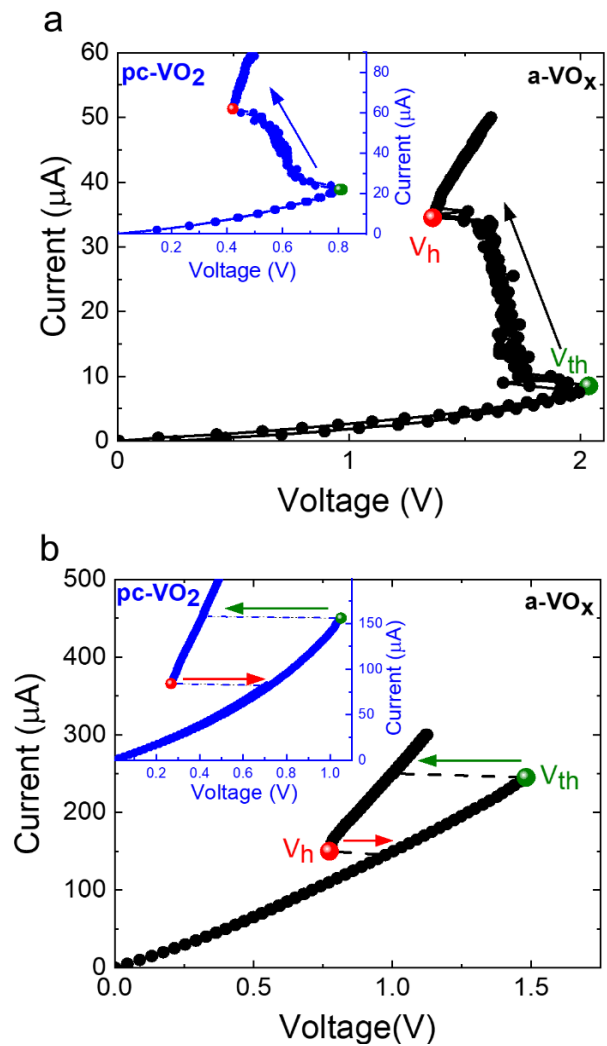

Figure 4: (a) S-type NDR in an a-VO$_x$ capacitor structure and inset showing similar behaviour in a $2\mu m \times 2\mu m$ pc-VO$_2$ cross-point device. (b) Snap-back NDR characteristics an a-VO$_x$ capacitor structure and inset showing similar behaviour in a $5\mu m \times 5\mu m$ pc-VO$_2$ cross-point device.

Of particular interest in this study is the fact that these devices can also exhibit a discontinuous 'snap-back' characteristic, such as that shown in Figure 4b. This is an alternative switching mode characterised by an abrupt increase in conductivity as the current reaches its threshold value and an abrupt reduction in conductivity as the current returns to its hold value. For the cases shown in Figure 4b, this hysteretic snap-back mode was effected by increasing the device area of the pc-VO$_2$ cross-point device from 4 μm$^2$ to 25 μm$^2$, and by increasing the electroforming compliance current for the a-VO$_x$ capacitor structure. In both cases, the transition is associated with a reduction in the device resistance but this alone does not explain the origin of snap-back response.

### 3.3 Temperature dependence

To gain further insight into these switching modes I-V characteristics were also investigated as a function of temperature, and Figure 5 shows typical results. The subthreshold I-V characteristics of both pc-VO$_2$ and a-VO$_x$ devices are well modelled by a trap-limited conduction model (e.g. Poole-Frenkel conduction[20]), while voltage controlled threshold switching and current-controlled snap-back characteristics of a-VO$_x$ devices serve to illustrate the systematic reduction of the threshold voltages ($V_{th}$) and hold voltages ($V_h$) with increasing temperature. The inset in Figure 5b also highlights the presence of discrete resistance changes during the metal to insulator transition, as previously reported for both thermal and voltage cycling of VO$_2$ devices where it was attributed to the heterogeneous nature of the transition[11,21].

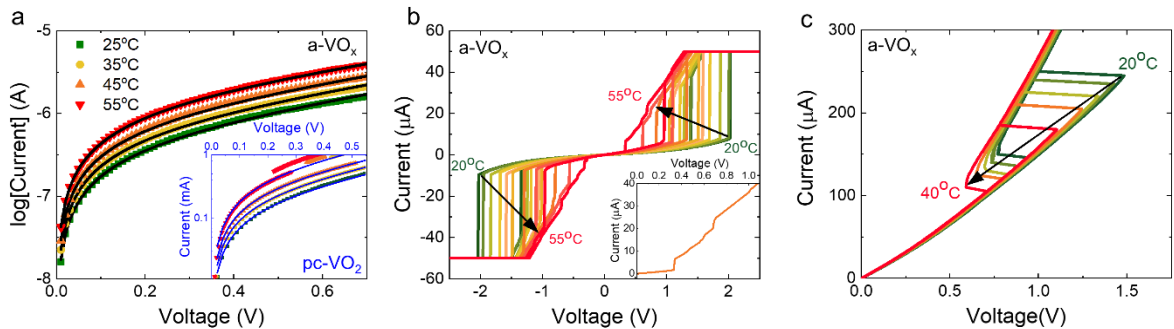

Figure 5: (a) Sub-threshold I-V characteristics of a post-formed a-VO$_x$ capacitor structure as a function of temperature, and inset showing similar behaviour for a 5μ × 5μ pc-VO$_2$ cross-point device; (b) voltage-controlled threshold switching response of an a-VO$_x$ device as a function of temperature; and (c) Current-controlled NDR response of an a-VO$_x$ device as a function of temperature.

Results from such measurements are summarized in Figure 6 which shows the temperature dependence of $V_{th}$ and an Arrhenius plot of the sub-threshold device resistance for selected pc-VO$_2$ and a-VO$_x$ devices. For switching based on a thermally induced metal-insulator transition, V$_{th}$ is expected to decrease with increasing device temperature ($T_0$) due to the fact that the transition temperature ($T_{IMT}$) can be achieved at lower power[22, 23]. In this case the filament temperature ($T$) can be approximated by a lumped element model such that $T = T_0 + R_{therm}IV$, where $R_{therm}$ is the thermal resistance of the filament, $I$ is the device current and $V$ is the device voltage. Using Ohm's law and a assuming a thermally activated device resistance of the form $R = R_0 e^{-E_a/kT}$ then reveals that $V_{th}^2 = \frac{R_0}{R_{therm}}(T_{IMT} - T_0)e^{E_a/kT_{IMT}}$. i.e. $V_{th}^2$ scales linearly with the device temperature $T_0$ and goes to zero as $T_0$ approaches $T_{IMT}$. As shown in Figure 6a, the measured V$_{th}$ for pc-VO$_2$ and a-VO$_x$ devices satisfies this equation and has an intercept in the range 340-350 K, consistent with the IMT in VO$_2$.[21, 24] Significantly, this temperature is similar for pc-VO$_2$ and a-VO$_x$ devices that exhibit continuous S-type NDR and for the a-VO$_x$ device that exhibits abrupt snap-back characteristics. This confirms the role of VO$_2$ in the switching of a-VO$_x$ devices and suggests that both the S-type and snap-back characteristics have a common origin.

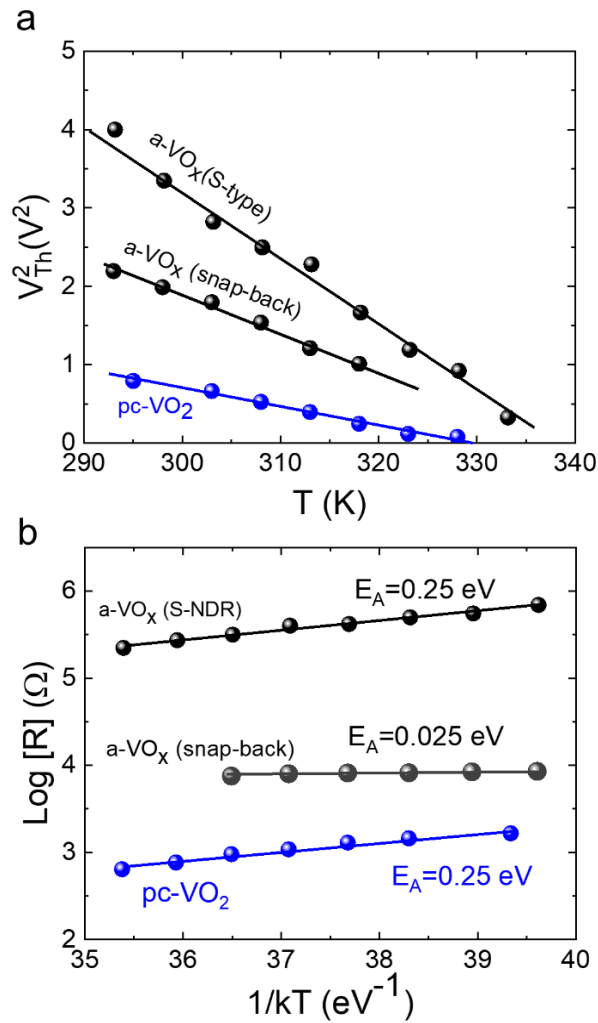

Figure 6: (a) Temperature dependence of the threshold voltage, and (b) an Arrhenius plot of the sub-threshold resistance for a-VO$_x$ devices that exhibit S-type and snap-back NDR and for a 5μm x 5μm pc-VO$_2$ device that exhibits snap-back NDR.

The associated Arrhenius plot shows that the change in sub-threshold resistance of the devices that exhibit S-type NDR is well characterised by a single activation energy of ~0.25 eV, consistent with previously reported values for conduction in polycrystalline VO$_2$[25, 26]. This further reinforces the view that VO$_2$ is the dominant phase both the a-VO$_x$ and pc-VO$_2$ devices. However, the sub-threshold resistance of the a-VO$_x$ device that exhibits a snap-response is characterised by an activation energy of ~0.025 eV, similar to that observed for the low resistance state of oxide-based resistive switching devices following an electroforming or a 'set' operation.[27] This is a particularly revealing as it suggests that subthreshold conduction is dominated by a high conductivity path through the oxide film even though the device exhibits a switching response characteristic of the IMT in VO$_2$.

### *3.4 Proposed model*

Both pc-VO$_2$ and a-VO$_x$ devices were shown to exhibit two distinct modes of CC-NDR, a smooth S-type mode or an abrupt snap-back mode, with the dominant switching mode dependence on the device area and the electroforming conditions. To understand this behaviour we draw on results from parallel studies in NbO$_x$-based devices where the NDR is attributed to the temperature dependence of trap-assisted conduction (e.g. Poole-Frenkel conduction[7, 12, 28]).

Using NbO$_x$-devices as a prototypic example of filamentary threshold switching we showed that the snap-back mode of CC-NDR can arise from a current redistribution process in which the current flowing in the region surrounding the conductive filament abruptly concentrates within the filament as it exhibits NDR in response to local Joule heating[7]. This can be understood by representing the device by a core-shell structure in which the core represents the high conductivity filament and the shell represents the parallel resistance due to conduction in the surrounding film. Simple circuit analysis then shows that current redistribution is controlled by the relative magnitudes of the NDR of the core, $R_{NDR}$, and the resistance of the shell, $R_s$, with continuous S-type characteristics observed for $R_s > R_{NDR}$ and abrupt snap-back characteristics observed for $R_s < R_{NDR}$[12]. This implies that the current in the surrounding area must be comparable to that in the filament in order to observe the snap back response and highlights the fact that the dominant behaviour will depend on the resistivity, thickness and area of the oxide film. Significantly, this analysis is independent of the process responsible for NDR.

In the present case, pc-VO$_2$ films were found to be much more conductive than a-VO$_x$ films and, as a consequence, the device resistance of 2 μm x 2 μm pc-VO$_2$ cross-point devices was comparable to that of 100 μm diameter a-VO$_x$ capacitor structures. In both cases these devices exhibited continuous S-type NDR when electroformed using low compliance currents. However, when the area of the pc-VO$_2$ cross-point devices was increased to 5 μm x 5 μm (a factor of 6.25 ) they exhibited a snap-back response, and when the 100 μm diameter a-VO$_x$ capacitor structures were electroformed using high compliance currents they also exhibited a snap-back response. Within the framework of the core-shell model the behaviour of the pc-VO$_2$ cross-point devices can be understood by accounting for the effect of the device area on the magnitude of the shell resistance, $R_s$. i.e. For the small area devices $R_s>R_{NDR}$ and the devices exhibit continuous S-type characteristics, while for the large area devices $R_s<R_{NDR}$, and they exhibit snap-back characteristics. Indeed, similar behaviour has previously been reported for NbO$_x$ devices.[5, 7, 12]

The effect of electroforming on the a-VO$_x$ devices requires further explanation. In this case, the device area was fixed and it is tempting to attribute the snap-back response to a change in $R_{NDR}$. However, we have found no clear correlation between $R_{NDR}$ and the electroforming conditions. Instead, we refer to the temperature dependent measurements in Figure 6 which show that sub-threshold conduction in a-VO$_x$ devices electroformed using high compliance currents is characterised by an activation energy of 0.025 eV, much lower than the 0.25 eV observed for devices formed at low compliance currents and for comparable pc-VO$_2$ devices. Despite this low activation energy, the NDR response of the devices remains consistent with the IMT in VO$_2$, suggesting that the filament consists of both highly conductive VO$_x$ and pc-VO$_2$. Given the nature of the electroforming process, the filament is expected to have radial symmetry and to consist of a central pc-VO$_2$ core and a halo of substoichiometric VO$_{x-\delta}$.[10, 29] The effect of electroforming can then be understood on the basis that the relative sizes of the pc-VO$_2$ core and VO$_{x-\delta}$ halo depend on the compliance current, with low compliance currents producing filaments that are dominated by pc-VO$_2$ and high compliance currents producing filaments with pc-VO$_2$ core and a significant VO$_{x-\delta}$ halo.

Figure 7 shows schematic representations of the proposed filament structures in a-VO$_x$ devices electroformed with 'low' and 'high' compliance currents, together with an equivalent electrical circuit. The relative diameters of the core and halo regions are assumed to increase with increasing forming current, consistent with the observed reduction in filament resistance and its temperature dependence. From an electrical perspective, the device can then be considered as three parallel resistors: one associated with the core and having a temperature dependent resistance governed by the heterogeneous IMT of VO$_2$; the second with the halo region and having a resistance determined by the electroforming conditions; and the third with the surrounding film and having a resistance determined by the stoichiometry, thickness and area of the oxide film. To a reasonable approximation the resistance of the halo and the surrounding film can be treated as constants so that the model reduces to the core-shell model discussed earlier, with the core represented by a temperature dependent resistor and the shell by a fixed parallel resistor of magnitude: $R_s = \frac{R_h R_f}{R_h + R_f}$, where $R_h$ is the effective resistance of the filament halo and $R_f$ the effective resistance of the surrounding film. Based on this model, the observation that the a-VO$_x$ devices exhibit S-type NDR at low compliance currents and snap-back NDR at high compliance currents can also be a attributed to a change in $R_s$, albeit from change in the halo resistance as a result of electroforming rather than a change in device area.

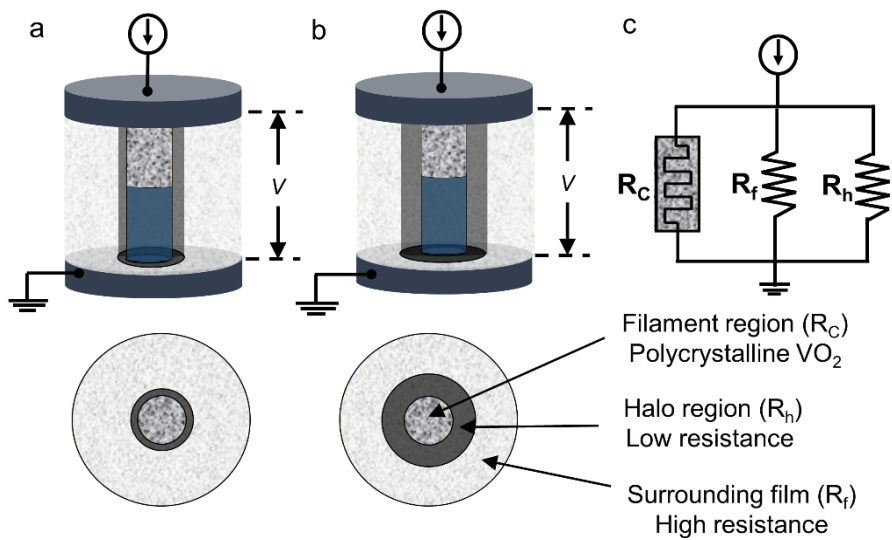

Figure 7: Schematic of the proposed filamentary core-shell structure produced by electroforming at (a) low and (b) high compliance currents and (c) the equivalent circuit model.

## 4. *Summary and Conclusions*

The voltage controlled threshold switching and current-controlled NDR behaviour of electroformed a-VO$_x$ and pc-VO$_2$ devices was investigated. Temperature dependent electrical measurements were consistent with the IMT in $VO_2$, and with the crystallisation of this phase within the a-$VO_x$ films during electroforming. Following electroforming each device type exhibited two distinct modes of NDR, a continuous S-type response and an abrupt snap-back response, depending on the device area and electroforming conditions. This behaviour was interpreted with reference to a simple core shell model of filamentary conduction in which the core represented the high-conductivity filament and the shell represented a parallel resistance due to the surrounding film. This predicted a transition between the S-type and snap-back behaviour based on the relative magnitudes of core NDR the shell resistance. For the pc-$VO_2$ devices the transition was demonstrated by using the device area to vary the shell resistance, as previously reported for a-$NbO_x$ devices. However, in the case of a-$VO_x$ devices it was controlled by the electroforming conditions. In that case, electroforming with high compliance currents was shown to produce composite filaments that were characterised by a high conductivity and weak temperature dependence at sub-threshold current but by the IMT of $VO_2$ at high currents. This was represented by a pc-$VO_2$ core and a surrounding substoichiometric halo, with the high conductivity halo acting to reduce the overall shell resistance and thereby control the transition between S-type and snap-back modes. Significantly, these results show that the snap-back characteristic is a generic response of systems that exhibit NDR

## 5. Acknowledgements

This work was partly funded by the Australian Research Council (ARC) and Varian Semiconductor Equipment/Applied Materials through an ARC Linkage Project Grant: LP150100693. We would like to acknowledge access to NCRIS facilities at the ACT node of the Australian National Fabrication Facility (ANFF) and the Australian Facility for Advanced ion-implantation Research (AFAiiR) and thank Dr. Tom Ratcliff for comments and feedback on the manuscript. We also like to thank James Murray for Raman analysis, and Shimul Kanti Nath for the assistance during device processing and discussion on switching behaviour. Caelb Estherby would like to thank Australian Government Research Training Program Scholarship. Sujan Kumar Das wants to thank Jahangirnagar University, Bangladesh for granting study leave to pursue PhD program at the Australian National University.

## 6. Data Availability

The data that support the findings of this study are available from the corresponding author upon reasonable request.

## 7. References

1. M. D. Pickett, G. Medeiros-Ribeiro and R. S. Williams, Nat. Mater. **12** (2), 114 (2013).
2. R. Tobe, M. S. Mian and K. Okimura, Journal of Applied Physics **127** (19), 195103 (2020).
3. C. Nauenheim, C. Kuegeler, A. Ruediger and R. Waser, Applied physics letters **96** (12), 122902 (2010).
4. D. Morgan, M. Howes, R. Pollard and D. Waters, Thin Solid Films **15** (1), 123-131 (1973).
5. S. K. Nandi, S. K. Nath, A. E. El-Helou, S. Li, T. Ratcliff, M. Uenuma, P. E. Raad and R. G. Elliman, ACS Appl. Mater. Interfaces <https://doi.org/10.1021/acsami.9b20252> (2020).
6. C. Berglund, IEEE Transactions on Electron Devices **16** (5), 432-437 (1969).
7. S. K. Nandi, S. K. Nath, A. E. El-Helou, S. Li, X. Liu, P. E. Raad and R. G. Elliman, Adv. Funct. Mater. **29** (50), 1906731 (2019).
8. J. M. Goodwill and M. Skowronski, Journal of Applied Physics **126** (3), 035108 (2019).
9. X. Zhou, D. Gu, Y. Li, H. Qin, Y. Jiang and J. Xu, Nanoscale **11** (45), 22070-22078 (2019).
10. L. Shuai, L. Xinjun, N. Sanjoy Kumar and E. Robert Glen, Nanotechnology **29** (37), 375705 (2018).
11. Y. Sharma, J. Balachandran, C. Sohn, J. T. Krogel, P. Ganesh, L. Collins, A. V. Ievlev, Q. Li, X. Gao and N. Balke, ACS nano **12** (7), 7159-7166 (2018).
12. S. Li, X. Liu, S. K. Nandi, S. K. Nath and R. Elliman, Adv. Funct. Mater. (1905060) (2019).
13. S. Kumar, J. P. Strachan and R. S. Williams, Nature **548** (7667), 318 (2017).
14. S. K. Nath, S. K. Nandi, S. Li and R. G. Elliman, Appl. Phys. Lett. **114** (6), 062901 (2019).
15. M. Vos, X. Liu, P. Grande, S. Nandi, D. Venkatachalam and R. Elliman, Nucl. Instrum. Methods Phys. Res., Sect. B **340**, 58-62 (2014).
16. M. S. Mian, K. Okimura and J. Sakai, Journal of Applied Physics **117** (21), 215305 (2015).
17. G. Seo, B.-J. Kim, H.-T. Kim and Y. W. Lee, Current Applied Physics **14** (9), 1251-1256 (2014).
18. D. Li, A. A. Sharma, N. Shukla, H. Paik, J. M. Goodwill, S. Datta, D. G. Schlom, J. A. Bain and M. Skowronski, Nanotechnology **28** (40), 405201 (2017).
19. G. A. Gibson, Advanced Functional Materials **28** (22), 1704175 (2018).
20. D. Ielmini and Y. Zhang, Journal of Applied Physics **102** (5), 054517 (2007).
21. J. A. J. Rupp, M. Querré, A. Kindsmüller, M.-P. Besland, E. Janod, R. Dittmann, R. Waser and D. J. Wouters, Journal of Applied Physics **123** (4), 044502 (2018).
22. F. Chudnovskii, Zhurnal Tekhnicheskoi Fiziki **45**, 1561-1583 (1975).
23. A. Bugaev, B. Zakharchenia and F. Chudnovskii, Leningrad Izdatel Nauka (1979).
24. K. Okimura, N. Hanis Azhan, T. Hajiri, S.-i. Kimura, M. Zaghrioui and J. Sakai, Journal of Applied Physics **115** (15), 153501 (2014).
25. V. A. Blagojević, N. Obradović, N. Cvjetićanin and D. Minić, Science of Sintering **45** (3), 305-311 (2013).
26. M. J. Tadjer, V. D. Wheeler, B. P. Downey, Z. R. Robinson, D. J. Meyer, C. R. Eddy Jr and F. J. Kub, Solid-State Electronics **136**, 30-35 (2017).
27. X. Liu, S. K. Nandi, D. K. Venkatachalam, S. Li, K. Belay and R. G. Elliman, presented at the 2014 Conference on Optoelectronic and Microelectronic Materials & Devices, 2014 (unpublished).

28. S. Slesazeck, H. Mähne, H. Wylezich, A. Wachowiak, J. Radhakrishnan, A. Ascoli, R. Tetzlaff and T. Mikolajick, RSC Adv. **5** (124), 102318-102322 (2015).
29. Y. Ma, J. M. Goodwill, D. Li, D. A. Cullen, J. D. Poplawsky, K. L. More, J. A. Bain and M. Skowronski, Adv. Electron. Mater. **5** (7), 1800954 (2019).